\documentclass[fleqn,twoside]{article}
\usepackage{espcrc2}
\usepackage{graphicx}
\usepackage{epsf}
\newcommand{\cd}{\makebox[0.08cm]{$\cdot$}}
\title{Two-fermion bound states in the explicitly covariant Light-Front
Dynamics}

\author{
V. A. Karmanov\address{Lebedev Physical Institute,\\ 
Leninsky Pr. 53, 119991 Moscow, Russia},
J. Carbonell\address{Institut des Sciences Nucl\'eaires, \\ 
        53, avenue des Martyrs, 38 026 Grenoble, France}
and     
M. Mangin-Brinet\addressmark\thanks{Now at D.P.N.C., University of Geneva/CERN}}

\begin{document}

%
%

\maketitle

\section{CONSTRUCTING THE ANGULAR MOMENTUM}

We solve the bound state problem for two relativistic fermions in the ladder
approximation.  The problem is
studied in the framework of the explicitly covariant light-front dynamics
\cite{karm76,cdkm}.  In this approach, the state vector is defined on an
hyperplane given  by the invariant equation $\omega\cd x=0$ with $\omega^2=0$. 
The standard light-front,  reviewed in \cite{BPP_PR_98}, is recovered for
$\omega=(1,0,0,-1)$.  Because of explicit covariance,  the wave functions for a
given angular momentum are decomposed in a few covariant spin structures
multiplied by scalar spin components. 

For two scalar particles with zero and non-zero angular momentum, this problem
was solved in \cite{These_MMB,2scalars}. The light-front deuteron wave function
was calculated, in a perturbative way, in \cite{ckj0}.  Now we develop a
general method, based on explicitly covariant light-front dynamics, to solve
the two-fermion bound state problem. We calculate the spin components 
non-perturbatively for one-boson exchange kernels: scalar, pseudoscalar, vector
and pseudovector exchanges \cite{These_MMB,mck_prd2,fermions}. 

The interaction kernel $V$ is a scalar, i.e., in c.m.-system, it depends on
the scalar products of all available vectors, including the vector 
$\vec{n}=\vec{\omega}/\vert\vec{\omega}\vert$, determining the light-front
orientaion, that is: $V=V(\vec{k},\vec{k'},\vec{s}_1,\vec{s}_2,\vec{n},M^2)$. 
Here $\vec{s}_1,\vec{s}_2$ are fermion spin operators. Thus, for the
two-fermion system, the angular momentum operator commuting with the kernel 
 reads:
\begin{equation}\label{ac3}                                                     
\vec{J}
=                                                                       
-i[\vec{k}\times \partial/\partial\vec{k}\,]
-i[\vec{n}\times                   
\partial/\partial\vec{n}]+\vec{s}_1+\vec{s}_2.                               
\end{equation}    
Besides, the three operators $\vec{J}^2, J_z$ and $A^2=(\vec{n}\cd\vec{J})^2$
commute not only with the kernel, but also with each other.  Therefore,  one
should construct the eigenstates of $\vec{J}^2,A^2$ (and $J_z$): 
\begin{eqnarray}\label{equ1}
\vec{J}^2\psi&=&J(J+1)\psi \\ A^2\psi&=&a^2\psi  
\label{equ2}  
\end{eqnarray} 
and solve the bound state problem for given ($J,a$). The equations for the
functions  with different ($J,a$) are not coupled with each other. Then for given
$J\geq 1$, one should construct from the states with different
$a$, a superposition satisfying the angular condition.

\subsection{The state with $J=0^+$}\label{sJ0}

The basis for the state with $J=0^+$ contains two spin structures. The counting
rule is based on the dimension of the spin matrix  forming the two-fermion wave
function: $N=(2s_1+1)(2s_2+1)/2=2\times 2/2=2$.  The division by 2 takes into
account the parity conservation. The vector $\vec{n}$
 enters in the wave function on the equal ground with
the momenta of particles.  In the reference system where
$\vec{k}_1+\vec{k}_2=0$, the  wave function of two fermions in the state $J=0^+$
is
represented as:
\begin{eqnarray}\label{eq0} 
{\mit \Phi}_{\sigma_2\sigma_1}&=&\sqrt{m}
w_{\sigma_2}^{\dagger}\psi(\vec{k},\vec{n})\sigma_y w_{\sigma_1}^{\dagger},
\nonumber\\
\psi(\vec{k},\vec{n})&=&
\frac{1}{\sqrt{2}}
\left(f_1+ \frac{i\vec{\sigma}\cd [\hat{k}\times \vec{n}]}
{\sin\theta}f_2\right),
\end{eqnarray}
where $\vec{k}=\vec{k}_1=-\vec{k}_2$, $\hat{k}=\vec{k}/k$. The functions
$f_{1,2}$ depend on $k$ and $\theta$, where $\hat{k}\cd\hat{n}=\cos\theta$. 
It is convenient to represent this wave function in the
four-dimensional form, in order to use in calculations the trace techniques of
the Dirac matrices. Namely,
\begin{equation}\label{eq1} 
{\mit \Phi}_{\sigma_2\sigma_1}(k_1,k_2,p,\omega\tau)=
\sqrt{m}\bar{u}_{\sigma_2}(k_2)
\phi U_c \bar{u}_{\sigma_1}(k_1)
\end{equation}
with
$
\phi=f_1S_1 +  f_2 S_2
$
and
\begin{eqnarray}\label{eq1_2} 
S_1&=&\frac{1}{2\sqrt{2}\varepsilon_k}\gamma_5,
\nonumber\\
S_2&=&\frac{\varepsilon_k}{2\sqrt{2} mk\sin\theta}
\left(\frac{2m\hat{\omega}}{\omega\cd p}- 
\frac{m^2}{\varepsilon_k^2}\right)\gamma_5,
\end{eqnarray}
where $\hat{\omega}=\omega_\mu\gamma^\mu$,  $U_c=\gamma^2 \gamma^0$, 
$\varepsilon_{k} = \sqrt{k^2+m^2}$. When $\vec{k}_1+\vec{k}_2=0$, eqs.
(\ref{eq1}), (\ref{eq1_2}) turn into (\ref{eq0}). The basis (\ref{eq1_2}) is
orthonormalized.

\subsection{The states with $J=1^+$}

The wave function of the $J=1^+$ state (deuteron, for example) is determined by
six spin structures. They are split in two families with two spin structures
for $J=1^+,a=0$ and with four spin structures for $J=1^+,a=1$.

One can easily check that the function $\vec{\psi}^0(\vec{k},\vec{n})$ 
satisfying the equation (\ref{equ2}) with $a=0$ is parallel to $\vec{n}$, i.e.,
it satisfies the condition  $\vec{\psi}^0=\vec{n}(\vec{n}\cd \vec{\psi}^0)$. It
has the following decomposition: 
\begin{eqnarray}\label{eq4a} 
\vec{\psi}^0(\vec{k},\vec{n})&=&\sqrt{\frac{3}{2}}\left\{g^{(0)}_1
\vec{\sigma}\cd\hat{k}\right. 
\nonumber\\ 
&+& \left. g^{(0)}_2
\frac{\vec{\sigma}\cd(\hat{k}\cos\theta-\vec{n})}
{\sin\theta}\right\}\vec{n}. 
\end{eqnarray} 
The function $\vec{\psi}^1(\vec{k},\vec{n})$ corresponding to 
$a=1$ is orthogonal to $\vec{n}$, i.e., it satisfies the condition 
$(\vec{n}\cd \vec{\psi}^1)=0$. 
It is convenient to introduce the vectors orthogonal to $\vec{n}$:
$$
\hat{k}_\perp= \frac{\hat{k}-\cos\theta\vec{n}}{\sin\theta},
\quad
\vec{\sigma}_\perp= \vec{\sigma}-(\vec{n}\cd \vec{\sigma})\vec{n}.
$$
Then the function $\vec{\psi}^1$ obtains the following form:
\begin{eqnarray}\label{eq4b} 
\vec{\psi}^1(\vec{k},\vec{n})&=&
g^{(1)}_1\frac{\sqrt{3}}{2}\vec{\sigma}_\perp
\nonumber\\
&+&g^{(1)}_2\frac{\sqrt{3}}{2}\left(2\hat{k}_\perp 
(\hat{k}_\perp \cd \vec{\sigma}_\perp)-\vec{\sigma}_\perp\right)
\\
&+&g^{(1)}_3\sqrt{\frac{3}{2}}\hat{k}_\perp (\vec{\sigma}\cd \vec{n})
+g^{(1)}_4\sqrt{\frac{3}{2}}i[\hat{k}\times \vec{n}]
\nonumber
\end{eqnarray}
Both wave functions (\ref{eq4a}) and (\ref{eq4b}) can be  also represented  in
the four-dimensional form \cite{mck_prd2,fermions}, similarly to (\ref{eq1}).
The functions $f_{1,2}$  for $J=0$ and $g^{(0)}_{1,2}$ for $J=1^+,a=0$ states are
determined by a system of two equations. The functions  $g^{(1)}_{1-4}$ for
$J=1^+,a=1$ state are determined by a system of four equations. The equations are
given in  \cite{These_MMB,mck_prd2,fermions,mck_prd1}.

\section{SATISFYING THE ANGULAR CONDITION}

The angular momentum operator in light-front dynamics is a dynamical one. Its
equivalence to the
kinematical operator (in particular,  to the operator (\ref{ac3}) for two
fermions) is ensured by the so called angular condition (see \cite{cdkm}). This
condition cannot be satisfied for the state with fixed $a\neq 0$, its solution
is a superposition of the states with different $a$. Therefore, the  wave
function for $J=1^+$, satisfying the angular condition, is represented as:
\begin{equation}\label{equ3}
\vec{\psi}(\vec{k},\vec{n})=c_0\vec{\psi}^0(\vec{k},\vec{n})
+c_1\vec{\psi}^1(\vec{k},\vec{n})
\end{equation}
Since we don't solve the full field theory, but restrict ourselves by a
particular (OBE) interaction, the angular condition can be also imposed in an
approximate form.  One can require that the wave function 
$\vec{\psi}(\vec{k},\vec{n})$ at $\vec{k}=0$  does not depend on $\vec{n}$.
The coefficients
$c_0,c_1$ in (\ref{equ3}) are unambiguously determined by this requirement. 

The masses $M_a$ for $a=0,1$ are split, though they should be degenerate in the
exact solution.  In standard LFD, this split -- the mass dependence on the
angular momentum projection -- is the violation of the rotational invariance. 
Solution (\ref{equ3}) corresponds to  average mass $M^2=c_0^2M_0^2+
c_1^2M_1^2$. For two scalar particles with massless exchange, this formula gives 
results close to the Bethe-Salpeter solution and restores with high
accuracy the Coulombien degeneration \cite{2scalars}.

One can show that at $k=0$, the components $g^{(0)}_{1,2}$  of the wave
function $\vec{\psi}^0$, eq. (\ref{eq4a}), depend on $\theta$ as:
$g^{(0)}_1(k=0,\theta)=b_0\cos\theta, \quad g^{(0)}_2(k=0,\theta)=
-b_0\sin\theta, $ where $b_0$ is a constant.  In the wave function
$\vec{\psi}^1$, eq. (\ref{eq4b}), the component $g^{(1)}_{1}(k=0,\theta)=b_1$
does not depend on  $\theta$, whereas  $g^{(1)}_{2-4}(k=0,\theta)=0$. The
coefficients in the solution (\ref{equ3}) are expressed through $b_0,b_1$:
\begin{equation}\label{ph2}
c_0=\frac{b_1}{\sqrt{2b_0^2+b_1^2}},\quad
c_1=\frac{\sqrt{2}b_0}{\sqrt{2b_0^2+b_1^2}}.
\end{equation}
The wave function (\ref{equ3}) is now can be rewritten as:
\begin{eqnarray*}\label{nz8} 
&&\vec{\psi}(\vec{k},\vec{n}) = f_1\frac{1}{\sqrt{2}}\vec{\sigma} + 
f_2\frac{1}{2}(3\hat{k}(\hat{k}\cd\vec{\sigma}) 
-\vec{\sigma}) 
\nonumber\\
&&+ f_3\frac{1}{2}(3\vec{n}(\vec{n}\cd\vec{\sigma}) 
-\vec{\sigma}) 
\nonumber \\
&&+
f_4\frac{1}{2}(3\hat{k}(\vec{n}\cd\vec{\sigma}) + 
3\vec{n}(\hat{k}\cd\vec{\sigma}) - 2(\hat{k}\cd\vec{n})\vec{\sigma}) 
\nonumber        
\\ &&+ f_5i\sqrt{\frac{3}{2}}[\hat{k}\times \vec{n}] +              
f_6\frac{\sqrt{3}}{2}(\vec{n}(\hat{k}\cd\vec{\sigma})
-\hat{k} (\vec{n}\cd\vec{\sigma})).           
\end{eqnarray*} 
The six components $f_i$ 
are expressed  through $g^{(0)}_{1,2}$ and $g^{(1)}_{1-4}$. 

\section{NUMERICAL RESULTS}

For illustration, we give here the results of numerical calculation with the 
kernel corresponding to the exchange of a scalar meson with mass $\mu$. The form
factor in the fermion-meson vertex is not introduced,  but the calculation
is done on the finite interval $k\leq k_{max}$. Note that for $J=0^+$ and
for the coupling constant $\alpha=g^2/(4\pi)<\alpha_c\approx 3.72$ the solution
is stable at $k_{max}\to\infty$ \cite{mck_prd1}.  For $J=1^+$, $m=1$, $\mu=0.25$ 
$\alpha=1.18$, $B=2m-M=0.0501$, $k_{max}=10$,   the energy split between the
states $a=0$ and $a=1$ is rather small: 2\%, in comparison to 20\% for two
scalar particles  \cite{2scalars}. For two fermions interacting by 
pseudoscalar exchange, the energy split is much more 
significant: for $\alpha=60$, $B_{a=0}=0.103$ and $B_{a=1}=0.0494$, that
is more than a 100\% splitting in energy \cite{fermions}.

\begin{figure}[htbp]
\vspace{-0.5cm}
\begin{center}
\epsfxsize=7.5cm\epsfxsize=7.0cm\mbox{\epsffile{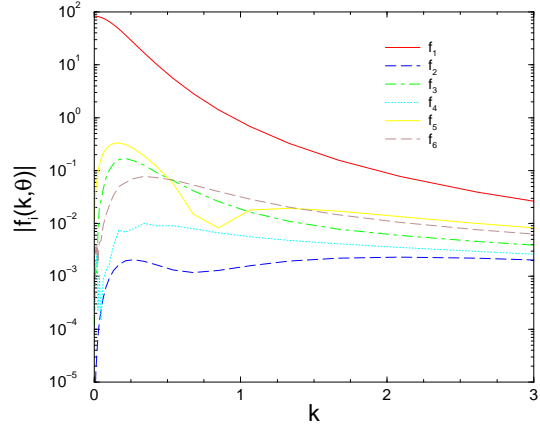}}
\end{center}
\vspace{-1.2cm}
\caption{The spin components $f_i(k,\theta=30^\circ)$ determining
the $J=1^+$ state wave function for scalar exchange.}
\label{fi} 
\vspace{-0.8cm}
\end{figure}

The components $f_i(k,\theta=30^\circ)$ are shown in figure \ref{fi}. One can
see that for scalar exchange the dominating component is $f_1$ (turning into
the S-wave in the nonrelativistic limit). For the coefficients (\ref{ph2}) the
following values were found: $c_0=0.5817,c_1=0.8134$, that is very close to
$1/\sqrt{3}\approx 0.5773$, $\sqrt{2/3} \approx 0.8165$. 

The details of approach and the numerical results for all exchanges (S, PS, V
and PV) will be published in \cite{fermions}.

\end{document}